# International research work experience of young females in physics

Serene H.-J. Choi, Maren Funk, Susan H. Roelofs, Martha B. Alvarez-Elizondo, and Timo A. Nieminen
The University of Queensland, Brisbane QLD 4072, Australia

*Abstract Summary*

*International research work for young people is common in physics. However, work experience and career plan of female workers in physics are little studied. We explore them by interviewing three international female workers in physics.*

I. INTRODUCTION

Typically, scientists, especially early in their careers, are highly and actively internationally mobile. Such mobility can be an important step in career development, in terms of the development of their professional academic learning and exposure to different life experiences.

The degree to which female workers in physics share this international mobility is little known. Physics is a scientific discipline with a notably low level of female participation, and this low level decreases with career progression—the female minority of research students becomes an even smaller minority among long-term research workers. Only a few female scientists are visible as outliers in science history [1]. On the other hand, a large part of the contributions by female scientists can be hidden or under-recognized, partly due to social and political pressures in patriarchal society [2]. Although the number of female scientists is small, their contribution to science is significant, and more significant than might be realized from the gallery of dead white men who populate the Physics Hall of Fame.

Since the early career period when the greatest international mobility occurs is also a time when the already small female participation in physics drops significantly, the experience of international mobility for females engaged in physics research is both interesting and of potential importance.

In our study, as an exploratory study, we investigate the perspectives and experiences of three young international female research workers about physics and research work through interview and discussion. The female research workers were from countries other than Australia, and were engaged in research work in physics at a research driven university in Australia.

II. METHODOLOGY

A. Participants

Three female research workers (designated interviewee 1, 2, and 3) in physics at an Australian university were individually interviewed. When interviewed, their residential durations in the host university varied from 2 and half months to 10 months on their limited visa status such as occupational trainee or working holiday maker. Two of them were postgraduate students and the last one has completed her Master's degree at their home. None of them had PhD degrees when interviewed. Their ages ranged from 23 to 28. Their nationalities are all different—countries in Western Europe and Central America. Interestingly, each of them already had past internship or study-abroad experience of various periods from 6 weeks to 6 months: One interviewee participated in one industry internship and two study abroad programs (studying physics, international politics, and economics) before. The other two interviewees also participated in study-abroad programs during undergraduate or high school periods.

B. Process

Five interviewees were invited to share notions on and experience in international research work as female workers in physics that is universally male dominated discipline. Three of them volunteered to participate in the interviews in order to publish the qualitative outcomes of the interviews as co-authors. The last two did not participate in this final stage due to busy schedules or privacy reasons. Involving interviewees by authorship is an effective qualitative method to get interviewee's insightful notions and collaboration by empowering them [3]. It also can be cost-effective due to the self-transcription by the interviewees instead of research assistants.

Before the interview, the interviewees filled out a questionnaire with a series of open-ended questions focusing on two topics: one, international work in host university at Australia; and another, own future career plan in physics. Then, semi-structured interviews [4] were individually conducted for about one hour to clarify and elaborate any issues in the questionnaire. If needed, further discussions including emails were used. The final interview data written in the questionnaire were thematically analysed [5].

III. RESULT AND DISCUSSIONS

A. *PART ONE: INTERNATIONAL WORK OR STUDY*

*1) Motivations for international work or study experience*

All the three female workers, in general, responded it was for more life experience in different culture and natural environment. They answered with a couple of motivational factors combined in their current and past international research work or study as below. No one factor was singled out.

- To explore future workplace (not only research but also industry). In the past multinational industry internship, for example, very good supervision and work atmosphere (e.g., regular presentations, applying theory to pragmatic applications, and lots interactions with other interns and people with various backgrounds) were experienced.
- Good to have it in CV.
- As a student without children and constraint employment, it is easy to go overseas and good to explore different cultures.
- Boyfriend or partner (working in different disciplines or industries) comes along.
- To take some courses that were not available at home institute.

On the basis of the motivational factors above, two female workers chose the Australian host institute to come to Australia or host city. The last chose it to join the research group that of her interest field for her PhD study.

*a) Workplace learning and experience*

The three female workers showed positive response in their technical research learning (technical aspects) and research cultural experience (supervision) during international work or study in current and past.

- In a past industry internship, interviewee 2 was encouraged by supervisor for a presentation and getting started in the lab (e.g., learning an experimental setup).
- At host institute in Australia, interviewee 3 had opportunities to learn experimental setup techniques and understand each of its components as well as measurement skills. Interviewee 2 reported a lot of freedom in doing experiments provided and initiative encouraged.
- Relationship with supervisors can be different even though they are in similar western higher education contexts. Interviewee 1 mentioned that she did not call the first names of lecturers in home institute compared to common calling the first names in Australia. She also commented about close supervision in host institute. When she had frequent direct interactions with supervisors or researchers in her research group (e.g., once a week) she was satisfied with it. Interviewee 3 experienced supervisor's excellent supports in term of open door policy (i.e., open to discuss with supervisors in any time) and research visits to other institutes if needed.

*b) Related life experiences in international research work or study in host institute at Australia*

All the three female workers found that people in Australian are easy going, talk friendly and willing to help (e.g., free riding to the hostel).

However, their experience with accommodation was different. While interviewee 2 had no difficulties to find accommodation within a week, interviewee 1 had difficulties in doing so. It seemed to depend on when they arrive. If their (often for short term staying) arrival is in the beginning of the semester, it is hard for them to find it because other domestic and international students (relatively for longer term staying) also are looking for accommodations.

One the issues of settle down quickly, the interviewees suggested more guidance with earlier information related to both research work that the intern would work on (e.g., related articles and reports) as well as local living facilities such as public transports, supermarkets, accommodations, and banking.

*2) Self reflection*

The female workers assessed their own achievements positively in their current and past international research work. They believed that employers would have positive impression about international experience in general because it shows independence and initiative.

All three female workers answered that they learned their fields of physics more and gained more experiences. In the past industry internship, the aim of the internship—a taste of working life (the goal of three month industry internship) was achieved by learning and experiencing a setup, a presentation, and writing a report (interviewee 2). In current work at Australia, more experience in experiments, journal papers and conference presentations were made (interviewee 3). More own independent lab work (e.g., nobody tell the intern what to do exactly) would enhance studies and writing thesis at home (interviewee 1). Also, some feasible advantages of host institute in Australia were commented such as improvement of English communication, and placed in high level of Times ranking that may give positive influence to future employer.

B. *PART TWO: PHYSICS AND MYSELF*

*1) Motivations for physics*

*a) Personal and social motivations of the female research workers*

All the three female workers answered own curiosity and desire to understand world and to know how things work that are happening around them were main motivation to study physics. They are interested in physics because it is challenging and provides with analytical skills. They believe that physics is associated with flexible professions, not necessarily in only academic research, but also process technology, consultancy, banking that requires analytical skills, developing environmental energy sources, and medical applications.

*b) Physics study in their schooling at home*

Unsurprisingly, all the three female workers answered that physics is not popular with girls at all at home. Most of their friends did not like physics because they considered it difficult. They pointed out rather own poor perception of females about physics subject can be problematic. Females might doubt themselves too much to do physics and think it is too difficult for them.

The female workers believed that curiosity, rather than brainy, would be more important characteristic to do physics. As a female worker in physics, they did not have particular female scientist idol. They have followed just own preference to be a physicist based on own interest in the subject. However, considering male dominancy in physics, one of the females commented that Marie Curie would be a symbol for the emancipation of women in science.

In schooling, their science teachers (for interviewee 1 and 2) and/or parents (interviewee 1 and 3) positively influenced or supported the female workers to study science subjects. Physics teacher provided them with positive comments on their curiosities and good marks. Parents discussed with them on scientific matters.

The female workers reported that being the only female student in the class at university is not an advantage because you are always noticed and do not like being the centre of attention. Or, they felt isolated because they were the only female in the group. When belonged to a group with mixed genders they feel much integrated because they could be involved in the social activities such as lunch together and laugh in workplace. It was better atmosphere for them to work.

*2) Future career plan*

Interestingly, two of the female workers hoped to get industry job related physics. In relation to any possibility of PhD study, interviewee 2 would consider any options to combine a PhD study with work in industry because she can apply physics to more immediate applications. Interview 1 would take PhD as most other students in physics at home. After PhD, she wanted to take an industry job because she wants to have experience more in society. The last wanted to get academic position.

All three of them showed their positive research work skills and reasonable confidence in their fields. They reviewed that own ability to be a physicist is good in logical thinking, creativity, and not backing off from difficulties, many scientific questions, and thinking of different approaches to solve a problem. For some certain topics, their enthusiasm is very high (individual research interests). They are willing to admit when they do not know and ask someone to advise.

The female workers did not believe any barriers for females to be professional physicists. When asked about any feasible prejudice or structural barriers for them to be a professional physicist, they were not sure because they have not experienced anything like that so far. However, they presumed that, in the future, if they have children it could be hard to keep up being as professional physicists. One of them commented that pregnancy leave should be managed not only for the female but also her partner.

In addition, they thought that it is good to have a female role model in their workplace. Having a female supervisor (with offspring) at the host institute in Australia, they could identify themselves with her and believe that it is feasible to be a professional physicist and a mother in their future. One of the females commented this (having female boss) was one of the reasons (with well-known research group of the field) for her to decide to come to the host university. She also mentioned that she needed to learn and develop the leadership skills (including determination and capabilities) shown by the female boss for herself in the future. That would attract more women in research.

### IV. LIMITATIONS

Having only three interviewees on their volunteerism in one institute, a caution has to be given not to overgeneralise the results. However, given only small number of females as minority group in physics in general, this may not be a critical issue. Indeed, other studies in investigating young female workers' perspective and experience in physics in international context have not found. Our qualitative interview may be the first exploratory study on the topic.

### V. CONCLUSION

The three females' perspective and experience at international work related physics are positive and not much different from those of young male research workers undertaking similar international research work (see [7]). Their learning is positive, progressive, and challenging in international research work. Based on their high motivation and sound preparation, they pursue their future careers in either industrial physics or academic.